# Measurement and Shaping of Circular Airy Vortex via Cross Phase

**Chen Wang (王 琛)[1], Yuan Ren (任 元)[1, 2]\***

[1]*Department of Aerospace Science and Technology, Space Engineering University,*

*Beijing 101416, China*

[2]*State Key Lab of Laser Propulsion & its Application, Space Engineering University,*

*Beijing 101416, China*

*\*Corresponding author: renyuan_823@aliyun.com;*

An optical vortex is a kind of structured light that possesses orbital angular momentum (OAM) per photon of $l\hbar$, where $l$ denotes topological charges (TCs) and $\hbar$ is Planck constant. As early as 1992, Allen showed that photons could carry OAM, which has aroused widespread concern among researchers[1]. Since then, optical vortices has been utilized in plethora of applications in the field of optical micro-manipulation[2], quantum entanglement[3] and rotation speed detection of objectives via the optical rotation doppler effect[4, 5]. Circular Airy beams have received intense interest due to their unique abruptly autofocusing property since they were realized in 2011[6, 7]. As a kind of optical vortices, the circular Airy vortex (CAV) was proposed since the spiral phase can improve some of the performance of circular Airy beams[8]. It's vital that we can measure the TCs and adjust the shape and even the singularity distribution of a CAV in the field of optical micro-manipulation. However, CAVs only appear at the self-focusing area, which makes it difficult to measure the TCs with conventional interference or diffraction methods[9, 10]. Besides, lots of methods have been proposed to achieve the shaping[11-14] or singularity manipulation of optical vortices[15, 16]. However, few methods can achieve these two functions simultaneously. Fortunately, we have the opportunity to achieve these goals at the same time with the cross-phase (CP).

In 2019, the CP, a new kind of phase structure, has been involved in Laguerre-Gauss (LG) beams that open up a new horizon for generation and measurement of optical vortices[9, 17]. Recently, we investigated a generation and measurement method of high-order optical vortices via the CP, which has been experimentally achieved[18]. In August 2020, we proposed a new type of CP, which can be employed to modulate optical vortices to implement shaping and multi-singularity manipulation simultaneously at far-field[19]. In October, inspired by the CP, we proposed a new kind of optical vortex called the Hermite–Gaussian-like optical vortex[20].

The form of the CP $\psi_0(x, y)$ in Cartesian coordinates $(x, y)$ is

$$\psi_0(x, y) = u(x^m \cos\theta - y^n \sin\theta)(x^m \sin\theta + y^n \cos\theta) \quad (1)$$

where the coefficient $u$ controls the conversion rate, the azimuth factor $\theta$ characterizes the rotation angle of converted beams in one certain plane. The order $n$ and $m$ are positive integers. When the two orders are both equal to 1 (the sum is 2), we call it low-order CP (LOCP); when the sum of these two orders is greater than 2, we call it high-order CP (HOCP). The LOCP can be engaged in the measurement of TCs; the HOCP could be used to implement shaping, even singularity manipulation of circular Airy. It is noteworthy that eq.(1) could be simplified to $\psi(x, y) = u x^m y^n$ when $\theta = 0$, and we only take this typical situation into count.

In this letter, we apply the CP to CAVs for the first time. On the one hand, we realize the measurement of TCs at the autofocusing area via the LOCP, on the other hand, we realize shaping intensity distributions to further improve the performance of CAVs with the HOCP. We also discuss the possibility of singularity manipulation of CAVs via the HOCP. In addition, there have been many reports about the generation of multi-singularity beams[21, 22], but lots of them are based on the astigmatic mode converter, which has high requirements on the accuracy, such as the relative position of the cylindrical lenses and oblique incidence angle, etc. Consequently, it's almost impossible to alter the positions of singularities precisely with the methods mentioned above. In contrast, the HOCP are more conducive to the precise manipulation of multi-particle and greatly evade the harsh requirements of the astigmatic mode converter with the help of a spatial light modulator (SLM), which is of great value in the field of multi-particle manipulation[20].

Without loss of generality, may the form of a CAV with the CP:

$$U(r, \phi, 0) = A_0 Ai(\Pi) \exp(a\Pi - \Pi^2 - il\phi) \psi(x, y) \quad (2)$$

where $A_0$ is the constant amplitude of the electric field, $Ai$ corresponds to the Airy function, $0 \le a < 1$ is the exponential truncation factor which determines the propagation distance, $l$ denotes TCs, and

$$\Pi = \frac{r_0 - r}{\omega} \qquad (3)$$

where $r_0$ denotes the radius of the primary Airy ring and $\omega$ is a scaling factor[23].

According to the Fresnel diffraction integral, when the light field mentioned above propagates a certain distance $z$, the output can be expressed that:

$$E(x,y,z) = \frac{1}{i\lambda z}\exp(ikz)\exp\left(\frac{ik}{2z}(x^2+y^2)\right) \\ \mathcal{F}\left[U(x_0,y_0,0)\exp\left(\frac{ik}{2z}(x_0^2+y_0^2)\right)\right] \qquad (4)$$

where $(x,y,z)$ denotes the observation plane, $\mathcal{F}$ is the Fourier transform.

Firstly, we would like to introduce the propagation properties of a CAV with the LOCP in the process of measuring TCs. Under the condition of the coefficient $\omega = 0.3\,\mathrm{mm}$, we simulate the propagation of the CAV with the LOCP from 0 to 0.25 $z/z_R$, as shown in Fig. 1. Akin to traditional CAVs, the propagation of the light field in Fig. 1(a) still has self-focusing properties. However, the light field distribution in the self-focusing area has a mode change due to the LOCP. This can be explained from two perspectives. The CAV with a LOCP is not an eigenmode, that is, it is not a set of solutions under the Helmholtz equation, so the stable propagation of its own light field cannot be guaranteed. On the other hand, the CP introduces astigmatism to a CAV, which makes the mode of the light field change during the propagation, as shown in Fig. 1(b) and (c). The LOCP is only employed to measure TCs of traditional optical vortices at far-field, while in the process of measuring CAVs the light field in the self-focusing area is a kind of Hermite–Gaussian-like mode, which can realize the measurement of TCs of CAVs. This method has two advantages. For one thing, compared with the far-field, this method can achieve the measurement of TCs in a short distance, saving space and optical components. For another thing, this method realizes the in-situ measurement of CAVs in the self-focusing area, which is meaningful in optical micro-manipulation.

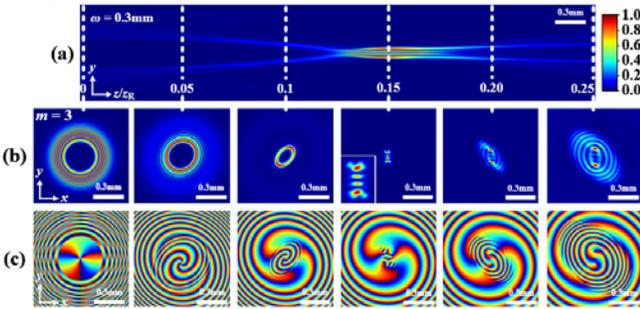

Fig. 1 The simulated propagation of the CAV with the LOCP in the process of measuring the TCs. (a) Side view of light intensity simulation in propagation in the Y-O-Z plane; the white dotted lines represent the propagation position, and the numbers in the dotted line represent the propagation distance. (b) Cross-sectional light intensity distributions corresponding to the white dashed lines in (a); the lower left corner is a partial enlarged view in the light intensity distribution at 0.15. (c) Phase distributions corresponding to (b).

The results of the measurement are shown in Fig. 2. The intensity distribution in the first column of Fig. 2(c) has two nodes, which can be deduced that the TCs of the original CAV are 3[18]. Further, the LOCP also can be adopted to measure the large TCs of CAVs, as shown in the third column of Fig. 2(c). After measuring, the TCs of the original CAV are 10. In addition, we can not only measure the value of TCs, but also measure the symbol of TCs, as shown in the second and fourth column of Fig. 2(c). The measured results of TCs are -3 and -10, respectively. The experimental results are shown in Fig. 2(d), which agrees well with the simulated results.

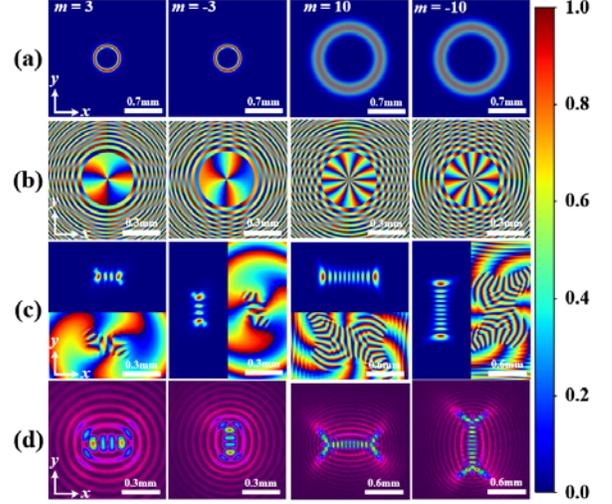

Fig. 2 Using the LOCP to realize the measurement of TCs of CAVs. (a) Simulated intensity distributions of CAVs with the LOCP at the initial plane. (b) Simulated phase distributions corresponding to (a). (c) Simulation distributions of intensity and phase in the self-focusing area. (d) Experimental intensity distributions corresponding to (c).

Secondly, we would like to introduce the propagation properties of a CAV with the HOCP in the process of shaping. Under the condition of the coefficient $\omega = 0.3\,\mathrm{mm}$, we simulate the propagation of the CAV with the HOCP from 0 to 0.5 $z/z_R$, as shown in Fig. 3(a). Bearing some analogy with traditional CAVs, the propagation of the light field in Fig. 3(a) still has self-focusing properties. In the self-focusing area, the CAV is shaped into a quadrilateral. Moreover, the distance of self-focus using the HOCP is significantly higher than that of the LOCP. This means that the use of the HOCP to achieve shaping of CAVs has a unique advantage that can perform beam shaping in the self-focusing area, arbitrarily adjusting the trajectory of the controlled particles by controlling the distribution of the energy flow. Further, under the condition of the coefficient $\omega = 0.5\,\mathrm{mm}$, we simulate the propagation of the CAV with the HOCP from 0 to 1 $z/z_R$, as shown in Fig. 3(b). In the part, we mainly focus on the changes of the CAV in the self-focusing area so we simulate the intensity distributions in the self-focusing area at the distance from 0.5 to 0.7 $z/z_R$, as shown in Fig. 3(b1) and (b2). The shaped CAV has a relatively stable distribution in the self-focusing area and a relatively long self-focusing distance, which is be of great importance for 3D particle manipulation. In addition, we can further increase the depth of focus (i.e. the self-focusing distance) by increasing the scale factor $\omega$. We can achieve beam shaping and focal depth control at the same time. This means that we can control the focus depth and trajectory of particles at the same time while we implement 3D particle manipulation.

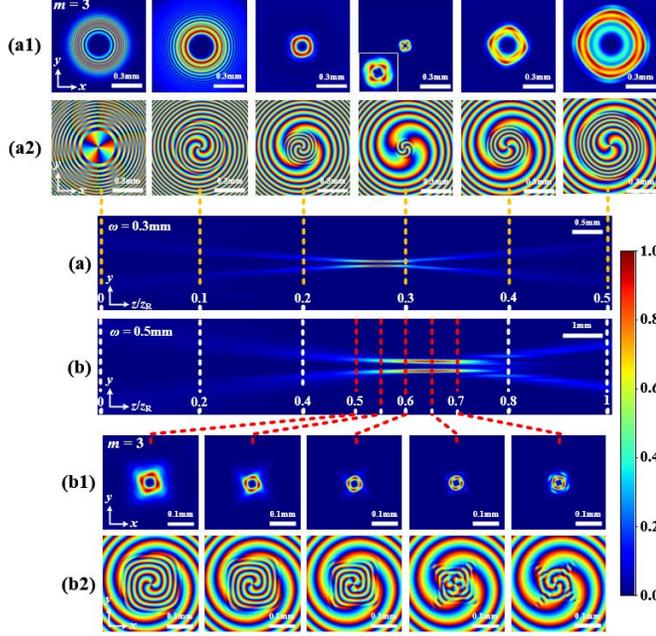

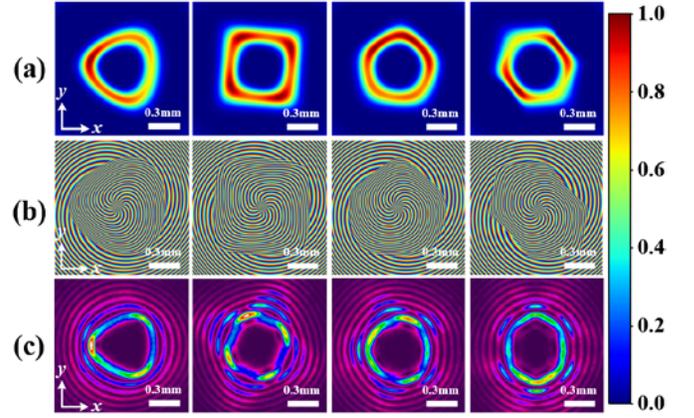

Fig. 3 The simulated propagation of the CAV with the HOCP in the process of shaping with different $\omega$. (a) Side view of light intensity simulation with $\omega=0.3$mm in propagation in the Y-O-Z plane; the yellow dotted lines represent the propagation position, and the numbers in the dotted line represent the propagation distance. (a1) Cross-sectional light intensity distributions corresponding to the yellow dashed lines in (a); the lower left corner is a partial enlarged view in the light intensity distribution at 0.3. (a2) Phase distributions corresponding to (a1). (b) Side view of light intensity simulation with $\omega=0.5$mm in propagation in the Y-O-Z plane; the yellow and red dotted lines represent the propagation position, and the numbers in the dotted line represent the propagation distance. (b1) Cross-sectional light intensity distributions corresponding to the red dashed lines in (b). (b2) Phase distributions corresponding to (b1).

Thirdly, we would like to introduce the polygonal shaping ability of the HOCP. The HOCP can not only shape the CAV into a quadrilateral shape, but theoretically shape the CAV into any symmetrical polygon. Interestingly, the number of sides of the shaped polygon is strictly equal to the orders of the HOCP. The orders of the HOCP are 3, 4, 5, 6, correspondingly. Consequently, the corresponding intensity distributions are triangle, quadrangle, pentagon and hexagon. Because manipulated particles move in the direction of energy flow, and changes in the intensity distribution can affect the distribution of OAM density, we can control the trajectory of the manipulated particles precisely by altering the order of the HOCP flexibly according to actual needs.

Fig. 4 Polygonal shaping of CAVs via the HOCP. (a) Simulated polygonal intensity distributions of CAVs shaped by the HOCP. The orders of the HOCP are 3, 4, 5, 6, correspondingly. (b) Simulated phase distributions corresponding to (a). (c) Experimental intensity distributions corresponding to (a).

Finally, we would like to discuss the possibility of singularity manipulation of CAVs via the HOCP. In fact, we can change the coefficient $u$ to adjust the degree of shaping. As shown in Fig. 5(b), by increasing the coefficient $u$, the degree of shaping gradually increases. Interestingly, the phase singularities of the light field are gradually splitting accordingly, as shown in Fig. 5(a). The white crosses denote the locations of singularities. This means that we can change the relative positions of singularities precisely by altering the parameter $u$, which is of great value in the field of multi-particle manipulation. Further, the singularities have been split into multiple singular points with $l=1$, which limits the ability of manipulation, but we can make up for this defect by increasing the incident optical power. In fact, the increase of the TCs means the increase of the radius of CAVs, and the particles orbit along the direction of the phase gradient of optical vortices, which limits the minimum distance in multi-particle manipulation.

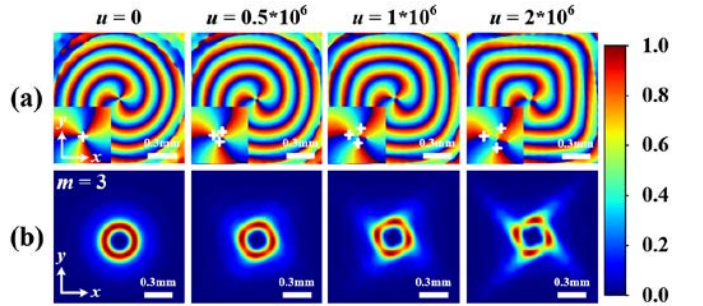

Fig. 5 Singularity manipulation of the CAV via the HOCP. (a) Simulated phase distributions of CAVs modulated by the HOCP. The white crosses denote the locations of singularities.

In summary, we apply the CP to CAVs for the first time experimentally. Firstly, we introduce the propagation properties of a CAV with the LOCP in the process of measuring TCs, and measure the TCs of CAVs in the self-focusing areas experimentally. Secondly, we shape the CAVs via the HOCP. Thirdly,

This work was supported in part by the National Nature Science Foundation of China under Grant

11772001 and 61805283, and in part by the Key Basic Research Projects of the Foundation Strengthening Plan under Grant 2019-JCJQ-XX-XXX.